\documentclass[prl,reprint,aps,showpacs]{revtex4-1}
\usepackage{graphicx}
\usepackage{subfigure}
\usepackage{dcolumn}
\usepackage{bm}
\usepackage{txfonts}
\usepackage{amssymb}

\newcommand{\bracket}[1]{\left ( #1 \right )}
\newcommand{\sqbracket}[1]{\left [ #1 \right ]}

\newcommand{\com}[2]{\left [ #1, #2 \right ]}
\newcommand{\acom}[2]{ \left \{ #1,#2 \right \}}

\newcommand{\Tr}[1]{{\rm Tr} \left( #1 \right)}
\newcommand{\aver}[1]{\left \langle #1 \right \rangle}

\begin{document}

\title{Dynamics of coupled vibration modes in a quantum non-linear mechanical resonator}

\author{G.~Labadze, M. Dukalski, and Ya.~M.~Blanter}
 \affiliation{Kavli Institute of Nanoscience, Delft University of
 Technology, Lorentzweg 1, 2628 CJ Delft, The Netherlands}

\date{20 August 2013}

\begin{abstract}
We investigate the behaviour of two non-linearly coupled flexural modes of a doubly-clamped suspended beam (nanomechanical resonator). One of the modes is externally driven. We demonstrate that classically, the behavior of the non-driven mode is reminiscent of that of a parametrically driven linear oscillator: It exhibits a threshold behavior, with the amplitude of this mode below the threshold being exactly zero. Quantum-mechanically, we were able to access the dynamics of this mode below the classical parametric threshold. We show that whereas the mean displacement of this mode is still zero, the mean squared displacement is finite and at the threshold corresponds to the occupation number of $1/2$. This finite displacement of the non-driven mode can serve as an experimentally verifiable quantum signature of quantum motion.
\end{abstract}

\maketitle

Observation of quantum effects in mechanical resonators, first reported in Ref. \onlinecite{Cleland2010} for a GHz resonator read out by a superconducting qubit, became a breakthrough in the field of nanomechanics. Subsequently, quantum effects were also confirmed in a mechanical drum resonator coupled to a superconducting microwave cavity \cite{Teufel2011} and in cavity optomechanical systems \cite{Painter2012,Kippenberg2012}. This breakthrough shifted the interest to the possible use of mechanical systems as quantum state transducers \cite{Didier,Palomaki} and eventually to the construction of integrated coherent mechanical-based circuits. Investigation of fundamental properties of coupled mechanical resonators is essential to achieve this goal.

Coupling of linear mechanical resonators or different modes of the same resonator has been extensively studied in the literature~\cite{Okamoto,Karabalin,Weig,Yamaguchi1,Yamaguchi2}. Recently, first experimental~\cite{Westra2010} and theoretical~\cite{Heikkila,Voje} studies of non-linearly coupled resonators were made available. They are facilitated by the fact that many available nanomechanical systems, such as suspended beams or membranes, are inherently non-linear due to elongated-induced stress. In the single-electron tunneling regime, non-linearities may be even stronger due to the Coulomb effects and can be controlled by nearby electric gates \cite{Usmani,Steele,Meerwaldt}. A basic property of a non-linear mechanical system is interaction between different vibrational modes. Stronger nonlinearity induces stronger coupling between these modes, which is highly beneficial for building integrated coherent circuits, classical as well as quantum ones.

Classically, non-linear systems exhibit extremely rich dynamical behavior, and in seemingly close situations they may behave very differently. Quantum effects in non-linear systems have been discussed in several contexts, including mechanical resonators \cite{Lifshitz-Cross}, and are generally recognized as a very complex and difficult problem. Non-linearity is essential for quantum position detection, since the mean expectation value of the displacement operator in a linear system is zero. The non-linear nature of a mechanical resonator can facilitate the transition into the quantum regime \cite{Katz2007}. In this Letter, we concentrate on one important aspect of quantum non-linear mechanical resonators, which is interaction between vibration modes.

We specifically consider a situation, when only one mode is externally driven. We first solve the classical problem and find that it is reminiscent of the parametrically driven oscillator, so that the non-driven mode only gets excited if the driving force exceeds certain threshold. Below the threshold, the classical displacement of the non-driven mode is exactly zero. Subsequently, we solve the quantum problem below the (classical) threshold using the Lindblad master equation technique and discover that quantum-mechanically, the non-driven mode gets excited to the states with non-zero number of phonons, up to the average occupation of one-half. This means that the occupation of the non-driven mode below the threshold is a quantum-mechanical effect and can serve as a signature of quantum motion. It also opens the way for detailed investigation of quantum dynamics of coupled mechanical oscillators such as for example entanglement generation or quantum state transfer between the modes.

\begin{figure}[b]
    \centering
    \includegraphics[width=.35\textwidth]{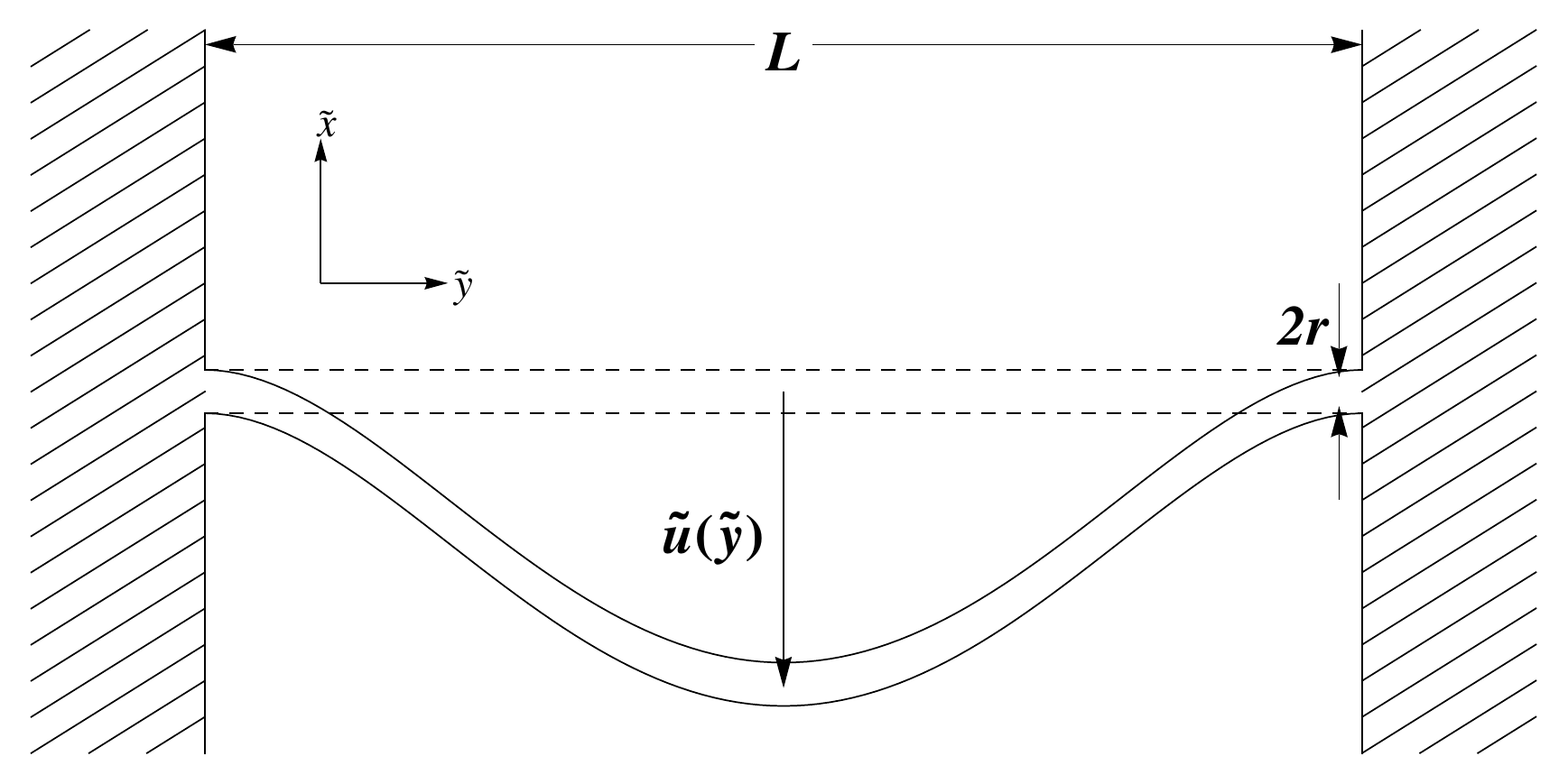}
    \caption{Schematic representation of a doubly clamped nanomechanical resonator of the length $L$ and the radius $r$. An applied force induces the bending profile $\tilde{u}(\tilde{y})$ as indicated.}
    \label{system}
\end{figure}

{\em Model}. To describe the interaction between the flexural modes of a doubly clamped nanomechanical beam, shown in Fig. \ref{system}, we use the Euler-Bernoulli equation \cite{LL}. We first derive the Hamiltonian of the beam. The latter is subject to the driving force $\tilde{F}\left(\tilde{t}\right)$, which can be of optical or magnetomotive in origin and induces the time-dependent bending profile $\tilde{u}(\tilde{y},\tilde{t})$. Displacement of the beam results in elongation which in turn induces the non-linear tension $\tilde{T}$. For simplicity, we use below the reduced coordinate $y = \tilde{y}/L$ along the beam and the reduced displacement $u = \tilde{u}/r$, where $L$ and $r$ are the length and the radius of the beam, which we take to be of a circular cross-section. Also we introduce the dimensionless time $t=\sqrt{D/\rho S L^4}\tilde{t}$, with $D$ being the bending rigidity, $\rho$ the mass density and $S$ the area of the cross section of the beam. The dimensionless tension $T=L^2\tilde{T}/D$  is given by Ref. \cite{LL},
\begin{eqnarray}
T&=&T_0+ \frac{K}{2} \int_0^1 dy \ \left (u''(y,t) \right)^2\,,
\end{eqnarray}
where $T_0$ is the residual tension of the beam and $K=r^2S/I$, with $I$ being the second moment of inertia. We denote by primes and dots spatial and temporal derivatives, respectively.
The applied force $\tilde{F}$, for which we use the dimensionless expression $F=L^4\tilde{F}/Dr$, can have static $F_{dc}$ and time dependent $F_{ac}$ components which result in dc and ac displacements of the beam $u(y,t)=u_{dc}(y)+u_{ac}\left(y,t\right)$ respectively. The equations of motion for these components have the form \cite{Clelandbook,Sampaz2003,Nayfeh1995}
\begin{eqnarray}
& & u''''_{dc} - T_{dc}u''_{dc} = F_{dc};  \label{Uacequationofmotion1} \\
& & \ddot{u}_{ac}+\eta\dot{u}_{ac}+\mathcal{L}[u_{ac}]-\bracket{T^*u''_{dc}+T_{ac}u''_{ac}}-T^*u''_{ac}=F_{ac}.
\label{Uacequationofmotion}
\end{eqnarray}
Here, $T_{dc}$ is the sum of the residual tension and the one resulting from the dc displacement; $T_{ac}$ is the tension term which contains all terms that are linear in $u_{ac}$, and $T^*$ is quadratic in $u_{ac}$. The operator $\mathcal{L}[u]$ is defined as
\begin{equation}
\mathcal{L}[u]=u''''-T_{dc}u''-T_{ac}[u]u''_{dc} \ .
\label{Loperator}
\end{equation}
The first three terms on the left-hand side of Eq. (\ref{Uacequationofmotion}) determine the linear response of the system. The last two terms introduce the nonlinearities with $u^2_{ac}$ and $u^3_{ac}$.

The eigenfunctions $\xi_n(y)$ and the eigenvalues $\omega_n$ of the operator $\mathcal{L}$ correspond to the mode shapes and frequencies of these modes respectively \cite{Nayfeh1995}. The ac displacement is expanded in terms of the mode shapes as $u_{ac}\left(y,t\right)=\sum^{\infty}_{n=1} \xi_n(y) u_n(t)/\sqrt{2\omega_n}$. Inserting this expansion in Eq. (\ref{Uacequationofmotion}) and taking the driving force to be a periodic function with the amplitude $F_0$ and the frequency $\omega_d$ provides a set of coupled equations of motion for the displacements of the modes, $u_n(t)$,
\begin{eqnarray}
\ddot{u}_n+\eta_n \dot{u_n}+\omega_n^2 u_n+\omega_n K \sum_{i,j}\bracket{2 A_i I_{nj}+ A_n I_{ij}}u_iu_j\nonumber\\
+\omega_n K \sum_{ijk}I_{ij}I_{kn}u_iu_ju_k=2\omega_n S_nF_0\cos(\omega_d t),
\label{acequationofmotion}
\end{eqnarray}
where the summation runs over all the eigenmodes, the values of $I_{ij}=\int \xi'_{i}(y)\xi'_j(y){\rm d}y$ depend only on the shapes of the modes $i$ and $j$, $S_n=\int \xi_n(y){\rm d}y$ is the mean displacement of the mode $n$ per unit deflection, and  $A_i=\int u'_{dc}(y)\xi'_i(y){\rm d}y$ depend on the static displacement. These coefficients can be calculated numerically. In the case of zero dc displacement, $A_i$ is zero, and the last term on the left-hand side of the Eq. (\ref{acequationofmotion}) couples the modes. Here we assume that the beam is in the strong bending regime where the dc displacement is big enough so that the geometrical nonlinearity plays an important role, but the time-dependent component of the deflection is small enough, and one can disregard nonlinearities which it causes. In this situation, we can disregard the last term in Eq. (\ref{acequationofmotion}). This statement imposes constraints on the ac displacement which can be found from the following inequality, $ \sum_{i,j}\bracket{2 A_i I_{nj}+ A_n I_{ij}}u_iu_j \gg\sum_{ijk}I_{ij}I_{kn}u_iu_ju_k$, see Ref. \onlinecite{Heikkila} for more details.

\begin{figure}[t]
    \hspace{.35mm}
    \includegraphics[width=.23\textwidth]{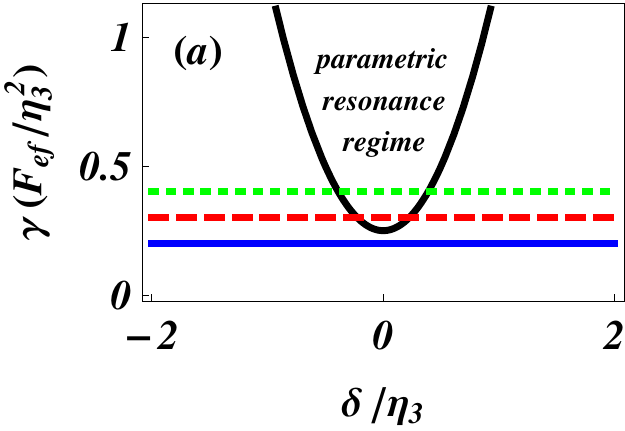}\label{parresreg}
    \hspace{4mm}
    \includegraphics[width=.22\textwidth]{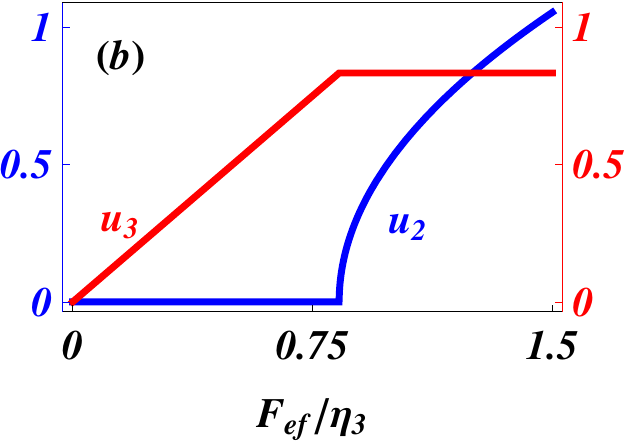}\protect\label{amplitudevsf}\\
    \includegraphics[width=.23\textwidth]{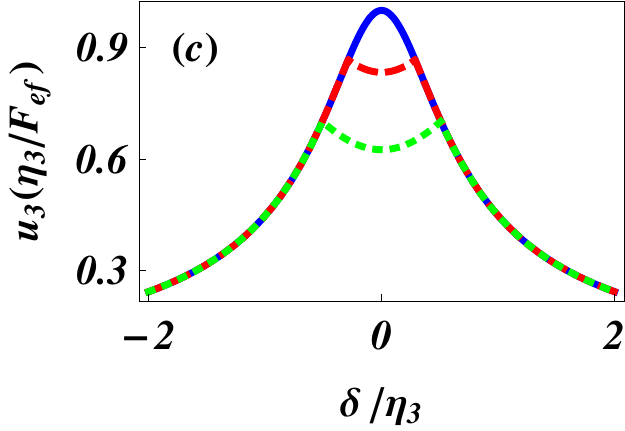}\protect\label{amplitude3vsdeltaA}
    \includegraphics[width=.235\textwidth]{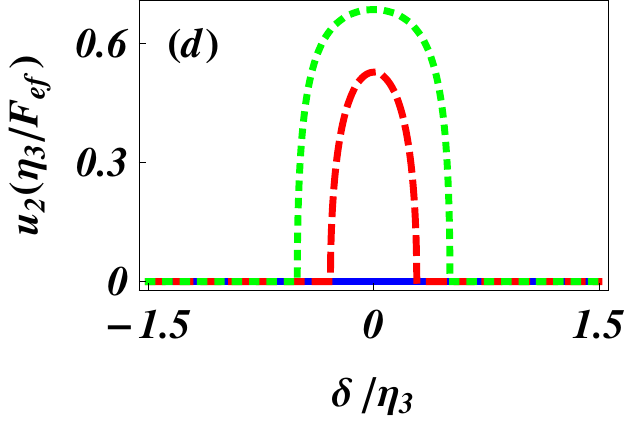}\protect\label{amplitude3vsdeltaB}
     \caption{(a) Stability diagram for the parametrically driven mode -- above the threshold (black line) the second mode is parametricaly driven by the third mode, $u_2 \ne 0$. The horizontal lines indicate the parameter values used to calculate the amplitudes of modes shown in (c,d). (b) The amplitude of the steady state of the coupled modes, the third one (red) and the second one (blue) at the resonance ($\delta=\Delta=0$) as a function of driving strength for fixed values of the coupling constant, $\gamma/\eta_3=0.3$ and the dissipation rate ratio, $\eta_2/\eta_3=1$. (c,d) Dependence of the amplitudes of the modes on the driving detuning $\delta$ for different values of the coupling constant $\gamma$ and fixed values of the dissipation rate ratio: $\eta_2/\eta_3=1$ and the intermode detuning $\Delta=0$.}
     \protect\label{amplitude3vsdelta}
\end{figure}

Note that since the external force is spatially homogeneous, it only can excite modes with odd $n$, for which $S_n \ne 0$. Our focus here is mode interaction, and therefore we only consider two modes, one of which is odd (driven), and another is even (not driven). Specifically, we take $n=2$ and $n=3$. This choice has an additional convenience since, as we show below, these modes are coupled the strongest in the quantum regime due to the frequency matching. For simplicity, we also disregard the Duffing terms (those with $i=j$ in Eq. (\ref{acequationofmotion}). They only renormalize the behavior of single modes, and this renormalization is well-known. The generalization of our theory to the case when these terms are present is straightforward. We are thus left with two coupled equations of motion,
\begin{eqnarray}
&& \ddot{u}_2+\eta_2\dot{u}_2+\omega_2^2u_2+4\omega_2\gamma u_2 u_3 = 0; \nonumber\\
&& \ddot{u}_3+\eta_3\dot{u}_3+\omega_3^2u_3+2\omega_3\gamma u_2^2 = 2\omega_3 F_{ef}\cos(\omega_d t),
\label{eqmo}
\end{eqnarray}
where $F_{ef}=S_3F_0$ and $\gamma=A_3I_{22}K/2$. Note that the intermode detuning $\Delta=2\omega_2-\omega_3$ is tunable by $u_{dc}$ and thus can be modulated by external gate voltage~\cite{Sampaz2003,Sazonova}. For simplicity, we assume that this quantity is small. This results in a stronger effective parametric coupling.

{\em Classical results}. Eqs. (\ref{eqmo}) describe two modes, one of which ($n=3$) is driven, and the other one ($n=2$) is parametrically coupled to former. We assume the the $n=3$ mode is driven close to its frequency. Eqs. \ref{eqmo} are easily solved in this resonance frequency approximation, which yields
\begin{eqnarray}
u_2&=&\left\{
                        \begin{array}{cl}
                        0& 4\gamma^2 F^2_{ef}<f;\\
                        \hspace{-3mm}\sqrt{\frac{2 \sqrt{4 F^2_{ef} \gamma ^2-\left(\frac{1}{2} \eta_3 (\delta +\Delta)+\delta\eta_2\right){}^2}+2 \delta  (\delta +\Delta )-\eta_2\eta_3}{2\gamma^2}}& 4\gamma^2 F^2_{ef}>f.
                        \end{array}
                \right.\\
 u_3&=&\left\{
                        \begin{array}{cl}
                        \frac{F_{ef}}{\sqrt{\delta^2+\bracket{\eta_3/2}^2}}& 4\gamma^2  F^2_{ef}<f;\\
                        \frac{\sqrt{(\delta +\Delta )^2+\eta_2^2}}{2 \gamma }& 4\gamma^2  F^2_{ef}>f.
                        \end{array}
                \right.\ \\
f &\equiv& \left[ \delta^2+\bracket{\eta_3/2}^2 \right] \left[ (\delta+\Delta)^2 +\eta_2^2 \right] \nonumber \ .
\end{eqnarray}
where $\delta=\omega_3-\omega_d \ll \omega_3$ is the detuning between the driving and the third mode frequencies.

For low values of the coupling constant $\gamma$, the amplitude of the third mode is not big enough to bring the second mode into the parametric resonance region, {\em i.e.} the effective coupling constant is below the parametric resonance threshold, $F_{ef} < \sqrt{f}/(2\gamma)$. Thus, in this case the second mode has zero amplitude while the third mode responds to the driving frequency in a simple harmonic manner, see Fig. \ref{amplitude3vsdelta}.
At the resonance and for sufficiently strong coupling $\gamma$, the system is driven over the threshold for $F_{ef} > \eta_2\eta_3/(4\gamma)$, so that the second mode has a finite amplitude, and the amplitude of the third mode stabilizes and does not depend on the force any more. The value of the threshold increases if one moves further away from the resonance.

{\em Quantum dynamics}.
Now we proceed with the quantum dynamics of the system of two interacting modes. First, we consider the dissipationless system. The starting point is the classical Hamiltonian,
\begin{eqnarray}
H=\int \limits_0^1 dy \left [ \frac{p^2}{2}+\frac 1 2 \mathcal{L}\sqbracket{u_{ac}} u_{ac}+\frac K 2 \bracket{\int_0^1u'_{dc}u'_{ac}dy}\bracket{u'_{ac}}^2 \right. \nonumber \\
\left.+\frac K 8 \bracket{\int_0^1\bracket{u'_{ac}}^2dy}\bracket{u'_{ac}}^2-\frac K 8 \bracket{\int_0^1\bracket{u'_{dc}}^2dy}\bracket{u'_{dc}}^2 \right ] \,
\end{eqnarray}
where we introduced the canonical momentum $p=\dot{u}_{ac}$. 
We promote both the position ${\hat u}_{ac}$ and the canonical momentum $\hat{p}$ to  operators acting within the Hilbert space, by imposing the standard commutation relations $\com{\hat{u}_{ac}(y)}{\hat{p}(y')}=i\delta(y-y')$. We can then write them in terms of creation and annihilation operators for phonons in individual modes,
\begin{eqnarray}
\hat{u}(y)&=&\sum_n \frac{\xi_n(y)}{\sqrt{2\omega_n}}\bracket{\hat{a}_n+\hat{ a}_n^\dagger}; \nonumber \\
\hat{p}(y)&=&-i\sum_n \sqrt{\frac{\omega_n}{2}}\xi_n(y)\bracket{\hat{a}_n-\hat{ a}_n^\dagger},
\label{sequ}
\end{eqnarray}
such that the operators $\hat{a}_n$ and $\hat{a}_n^\dagger$ obey the following commutation relations: $\com{\hat{a}_n}{\hat{a}_m^\dagger}=\delta_{nm}$ and zero otherwise.
Inserting Eq. (\ref{sequ}) in the Hamiltonian and using the orthogonality of eigenfunctions $\xi_n(y)$ yields
\begin{eqnarray}
\hat H_0&=&\sum_n \omega_n\bracket{\hat{a}_n^\dagger \hat{a}_n } \\
\hat H_I&=&\frac K 2 \sum_{kmn}A_k I_{mn} \bracket{\hat{a}_k+\hat{a}_k^\dagger}\bracket{\hat{a}_m+\hat{a}_m^\dagger}\bracket{\hat{a}_n+\hat{a}_n^\dagger}\nonumber\\
&+&\frac K 8 \sum_{ijmn}I_{ij}I_{mn}\bracket{\hat{a}_i+\hat{a}_i^\dagger}\bracket{\hat{a}_j+\hat{a}_j^\dagger}
\bracket{\hat{a}_m+\hat{a}_m^\dagger}\bracket{\hat{a}_n+\hat{a}_n^\dagger}. \nonumber
\end{eqnarray}
Again we assume that the static displacement is big enough to give rise to buckling induced  nonlinearities, but the time dependent ac deflection is small enough to disregard nonlinearities caused by it, $u_{dc} \gg u_{ac}$. In this regime one can drop the quartic term in the Hamiltonian.
We also assume that the system is driven by a classical field with the strength $F_0$ and the frequency $\omega_d$. The resulting Hamiltonian has the form $\hat H_{tot}=\hat H_0+\hat H_I+\hat H_{d}$, where
\begin{equation}
H_{d}=\sum_n S_n F_0\cos(\omega_d t)(\hat{a}_n+\hat{a}_n^\dagger)\,.
\end{equation}
Similarly to the classical case, we restrict ourselves to the second and third mode, assuming that the system is driven close to the frequency of third mode, $\omega_d\approx\omega_3$, and the intermode frequency detuning $\Delta=2\omega_2-\omega_3$ is small. Then the interaction picture Hamiltonian $\hat{\mathcal{V}}=\exp\left(-i \hat H_0 t\right)\left(\hat H_d+\hat H_I \right)\exp\left(i \hat H_0 t\right)$ has the form
\begin{eqnarray}
\hat{\mathcal{V}}&=&\frac{F_{ef}}{2}\bracket{e^{-i \delta t} \hat{a}_3+e^{i\delta t} \hat{a}_3^{\dagger}}+ \gamma \bracket{e^{i\Delta t}\hat{a}_3^{\dagger}\hat{a}_2\hat{a}_2+e^{-i\Delta t}\hat{a}_3\hat{a}_2^{\dagger}\hat{a}_2^{\dagger}} \nonumber \\
& \equiv &
\hat{\mathcal{V}}_D+\gamma \hat{\mathcal{V}}_I \ ,
\label{InteractionHamiltonian}
\end{eqnarray}
where all quickly rotating terms have been disregarded (rotating wave approximation).

\begin{figure}[t]
    \centering
     \includegraphics[width=.46\textwidth]{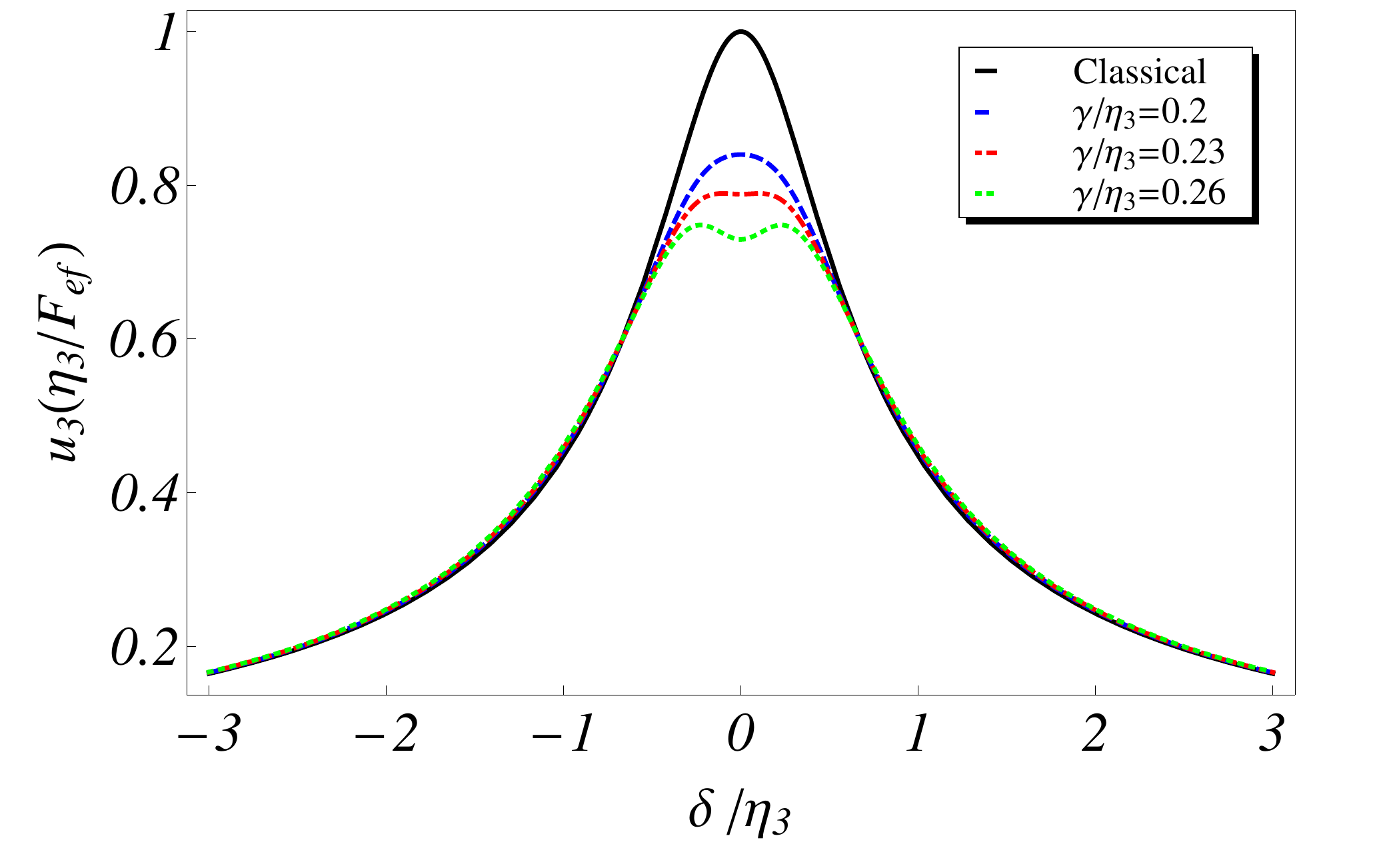}
     \caption{Dependence of the amplitude of the third mode on the driving detuning $\delta=\omega_3-\omega_d$ for different values of the coupling constant $\gamma$ and fixed values of the dissipation rate ratio, $\eta_2/\eta_3=1$, and the intermode detuning, $\Delta=2\omega_2-\omega_3=0$.}
     \label{u2quantumvsdelta}
\end{figure}

To include the effect of dissipation, we  model the evolution of the system with a Linblad type master equation at zero temperature,
\begin{equation}
\dot{\rho}_{tot}=\frac{1}{i}\com{\hat{\mathcal{V}}}{\rho_{tot}}+\eta_{2}\mathcal{D}\bracket{\rho_{tot},\hat{a}_2}+
\eta_{3}\mathcal{D}\bracket{\rho_{tot},\hat{a}_3},
\label{Lingblad}
\end{equation}
where $\eta_{2,3}$ are the dissipation rates of the two modes, and
$
\mathcal{D}\bracket{\rho,\hat{\Theta}} \equiv \hat{\Theta}\rho\hat{\Theta}^{\dagger}-
\frac{1}{2}\acom{\hat{\Theta}^{\dagger}\hat{\Theta}}{\rho}\,.
$
Here, $\rho_{tot}$ is the density matrix projected into the two mode subspace.
We justify the choice of this dissipation model by the fact that  a driven uncoupled quantum harmonic oscillator modelled with it shows  decay features, resembling those found in the classical description.
Furthermore, we do not expect that this will affect the qualitative features of our result in any way.

Individual modes can be described by the reduced density operators $\rho_2$ and $\rho_3$ acting on the correspondent subspace of the Hilbert space $\mathcal{H}$.  We note that here $\gamma$ will be the smallest parameter in our model, allowing us to study the problem perturbatively.

The condition that any physical density operator $\rho$ needs to be positive, with purity $\Tr{\rho^2}\leq 1$, imposes an upper-bound on the validity of  the applied perturbation approximation.
We numerically find that such perturbation is only valid within the parameter regime where the classical theory is below the (classical) parametric resonance threshold. Classically, in this region the third mode has the Lorentzian response while the second mode has no response (zero amplitude).

We solve  Eq. (\ref{Lingblad}) for the density matrices of the second and the third modes up to second order of the perturbation. The amplitude of the third mode $u_3$ reads
\begin{equation}
u_3=\frac{F_{ef}}{\sqrt{4\delta^2+\eta_3^2}}\left(1-\frac{8 \gamma^2 \left( \eta_2\eta_3-2\delta(\delta+\Delta) \right)}{\left((\delta+\Delta)^2+\eta_2^2\right)(4\delta^2+\eta_3^2)}\right)^{1/2}\,. \nonumber
\end{equation}
The dependence of the amplitude of the third mode on the driving frequency detuning is shown in Fig. \ref{u2quantumvsdelta}. One can see that in the quantum regime below the parametric resonance threshold the amplitude of this mode is reduced compared to the classical value. The amplitude can even acquire a double peak shape.
Based on our second order results, it is also easy to show that the state of the third mode stays coherent (Poissonian), and the Fano factor is equal to one, $F_3\equiv\Delta\hat{n}_3^2/\aver{\hat{n}_3}=1$.

\begin{figure}[t]
    \centering
    \vspace{3mm}
    \includegraphics[width=.35\textwidth]{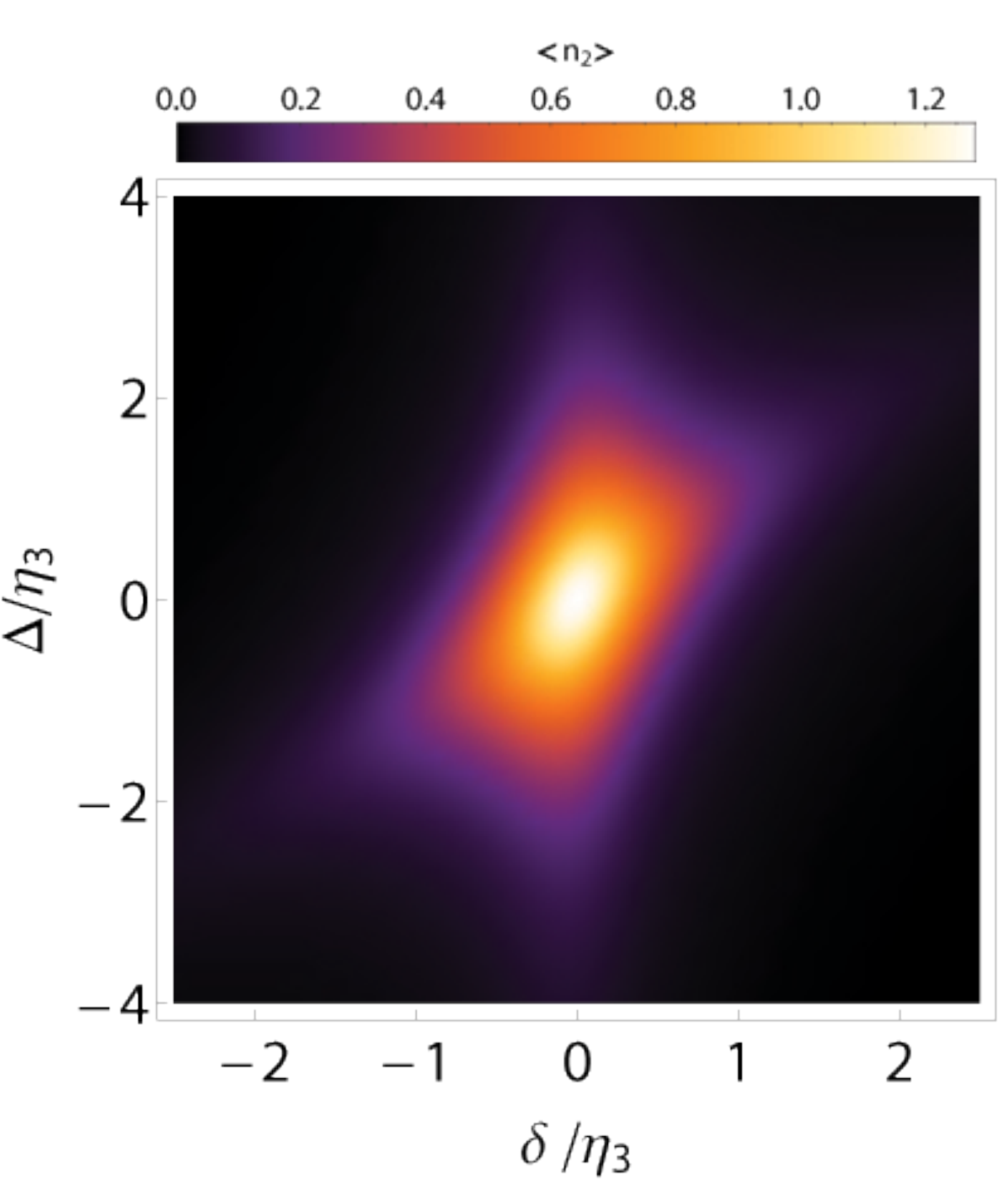}
    \vspace{-2mm}
    \caption{(a) Color plot of the average number of phonons in the second mode as a function of the intermode detuning $\Delta$ and the detuning from resonance frequency $\delta$.  The values of the dissipation rate ratio, the driving strength and the mode coupling were to: $\eta_2/\eta_3=1$, $F_{ef}/\eta_3=2$ and $\gamma/\eta_2=0.2$. }
    \label{Avern2plot}
\end{figure}

For the second mode in the quantum regime, the expectation value of displacement $\aver{\hat{u}_2}$ vanishes, however, the root-mean-square displacement does not due to a non-zero contribution to the  average number of phonons,
\begin{equation}
\aver{\hat{n}_2}=\frac{8 \gamma ^2 F_{ef}^2}{\left((\delta +\Delta )^2+\eta _2^2\right) \left(4 \delta ^2+\eta_3^2\right)}.
\end{equation}
Note that $\aver{\hat{n}_2} \leq 1/2$, with the value $1/2$ achieved exactly at the threshold. These phonons appear in addition to the usual zero-point motion, responsible for $1/2$ phonon independently of the driving force.

The resonance for this mode is observed in two cases, see Fig. \ref{Avern2plot}: when the driving frequency equal to the frequency of the third mode $\tilde{\omega}_d=\tilde{\omega}_3$ or when the system is driven at the double frequency of the second mode $\omega_d=2\omega_2$. The Fano factor for second mode is equal to $F_2\equiv\Delta\hat{n}_2^2/\aver{\hat{n}_2}=3/2$, which is an indication of bunching. This was expected due to the form of the coupling between the two modes (every time the particle is annihilated in the third mode two particles are created in the second mode) and the fact that phonons obey Bose-Einstein statistics, in combination with the decay mechanism which dissipates one excitation at the time the two-excitation bunched states produced by the intermodal coupling. To put the value of $3/2$ in perspective we note that for a coherent state (no bunching) the Fano factor is 1, however for bunched squeezed state (without dissipation) Fano factor is $2$ or higher.

Our main conclusion is that the non-driven mode below the classical parametric resonance threshold is not driven classically (the oscillation amplitude is zero), but is driven quantum-mechanically. Thus, the experimental observation of non-zero quanta of phonons in the non-driven mode below the threshold can serve as an indication that the mode is in the quantum regime. To arrive to this conclusion, we used a simplified model which (i) disregards non-linear nature of the resonator modes; (ii) disregards other vibrational modes. We do not expect the result to be qualitatively modified, however, quantitatively the inclusion of these factors is non-trivial and will be performed elsewhere.

The work was supported by the Dutch Foundation for Fundamental Research (FOM).


\begin{thebibliography}{99}

\bibitem{Cleland2010}  A. D. O\'{}Connell, M. Hofheinz, M. Ansmann, R. C. Bialczak, M. Lenander, E. Lucero, M. Neeley, D. Sank, H. Wang, M. Weides, J. Wenner,  J. M. Martinis, and A. N. Cleland, Nature {\bf 464}, 697 (2010).
\bibitem{Teufel2011}  J. D. Teufel,  D. Li, M. S. Allman, K. Cicak, A. J. Sirois, J. D.  Whittaker, and  R. W. Simmonds,  Nature {\bf 471}, 204 (2011).
\bibitem{Painter2012}   A. H. Safavi-Naeini, J. Chan, J. T. Hill,  T. P. M. Alegre, A. Krause, and O. Painter, Phys. Rev. Lett. {\bf 108}, 033602 (2012).
\bibitem{Kippenberg2012}  E. Verhagen, S. Del\'eglise, S. Weis, A. Schliesser, and T. J. Kippenberg, Nature {\bf 482}, 63 (2012).
\bibitem{Didier} N. Didier, S. Pugnetti, Ya. M. Blanter, and R. Fazio, Arxiv e-prints:1201.6293
\bibitem{Palomaki} T. A. Palomaki, J. W. Harlow, J. D. Teufel, R. W. Simmonds, and K. W. Lehnert, Nature {\bf 495}, 210 (2013).
\bibitem{Okamoto} H. Okamoto, T. Kamada, K. Onomitsu, I. Mahboob, and H. Yamaguchi, Appl. Phys. Express {\bf 2}, 062202 (2009).
\bibitem{Karabalin} R. B. Karabalin, M. C. Cross, and M. L. Roukes, Phys. Rev. B {\bf 79}, 165309 (2009).
\bibitem{Weig} T. Faust, J. Rieger, M. J. Seitner, P. Krenn, J. P. Kotthaus, and E. M. Weig, Phys. Rev. Lett {\bf 109}, 037205 (2012).
\bibitem{Yamaguchi1} H. Yamaguchi and I. Mahboob, New J. Phys. {\bf 15}, 015023 (2013).
\bibitem{Yamaguchi2} H. Okamoto, A. Gourgout, C.-Y. Chang, K. Onomitsu, I. Mahboob, E. Y. Chang, and H. Yamaguchi, Nature Physics {\bf 9}, 480 (2013)
\bibitem{Westra2010}  H. J. R. Westra,  M. Poot, H. S. J. van der Zant,  and  W. J. Venstra, Phys. Rev. Lett. {\bf 105}, 117205 (2010).
\bibitem{Heikkila} R. Khan, F. Massel, and T. T. Heikkil\"{a}, Phys. Rev. B {\bf 87}, 235406 (2013).
\bibitem{Voje} A. Voje, A. Isacsson, and A. Croy, Phys. Rev. A. {\bf 88}, 022309 (2013).
\bibitem{Usmani} O. Usmani, Ya. M. Blanter, and Yu. V. Nazarov, Phys. Rev. B {\bf 75}, 195312 (2007).
\bibitem{Steele} G. A. Steele, A. K. H\"uttel, B. Witkamp, M. Poot, H. B. Meerwaldt, L. P. Kouwenhoven, and H. S. J. van der Zant, Science {\bf 325}, 1103 (2009).
\bibitem{Meerwaldt} H. B. Meerwaldt, G. Labadze, B. H. Schneider, A. Taspinar, Ya. M. Blanter, H. S. J. van der Zant, and G. A. Steele,  Phys. Rev. B {\bf 86}, 115454 (2012).
\bibitem{Lifshitz-Cross} R. Lifshitz and M. C. Cross,  Review of Nonlinear Dynamics and Complexity {\bf 1}, 1 (2008).
\bibitem{Katz2007}  I. Katz, A. Retzker, R Straub, and R. Lifshitz, Phys. Rev. Lett. {\bf 99}, 040404 (2007).
\bibitem{LL} L. D. Landau and E. M. Lifshits, Theory of Elasticity (Butterworth-Heinemann, Burlington, 1986).
\bibitem{Clelandbook} A. N. Cleland, Foundations of Nanomechanics (Springer, Heidelberg, 2002).
\bibitem{Nayfeh1995}   A. H. Nayfeh, W. Kreider, and T. J. Anderson, AIAA Journal {\bf 33}, 1121 (1995).
\bibitem{Sampaz2003} S. Sapmaz, Y. M. Blanter, L. Gurevich, and H. S. J. van der Zant, Phys. Rev. B {\bf 67}, 235414, (2003).
\bibitem{Sazonova} V. Sazonova, Y. Yaish, H. Ustunel, D. Roundy, T. A. Arias, and P. L. McEuen, Nature {\bf 431}, 284 (2004).
\end{thebibliography}
\end{document}